\documentclass[aps,prl,twocolumn,superscriptaddress,showpacs]{revtex4}
\usepackage{graphicx}
\usepackage{dcolumn}
\usepackage{bm}

\begin{document}


\title{NMR evidence for the persistence of spin-superlattice above the 1/8 magnetization plateau in SrCu$_{2}$(BO$_{3}$)$_{2}$}

\author{M.~Takigawa}
\affiliation{Institute for Solid State Physics, University of Tokyo,
Kashiwanoha, Kashiwa, Chiba 277-8581, Japan}
\author{S.~Matsubara}
\altaffiliation[Present address: ]{Hitachi Central Research Laboratory, Kokubunji, Tokyo 185-8601, Japan}
\affiliation{Institute for Solid State Physics, University of Tokyo,
Kashiwanoha, Kashiwa, Chiba 277-8581, Japan}
\author{M.~Horvati\'{c}}
\affiliation{Grenoble High Magnetic Field Laboratory, CNRS, BP 166 - 38042
Grenoble, France}
\author{C.~Berthier}
\affiliation{Grenoble High Magnetic Field Laboratory, CNRS, BP 166 - 38042
Grenoble, France}
\author{H.~Kageyama}
\affiliation{Department of Chemistry, Graduate School of Science, Kyoto
University, Kyoto 606-8502, Japan}
\author{Y.~Ueda}
\affiliation{Institute for Solid State Physics, University of Tokyo,
Kashiwanoha, Kashiwa, Chiba 277-8581, Japan}

\date{\today}

\begin{abstract}
We present $^{11}$B NMR studies of the 2D frustrated dimer spin system
SrCu$_{2}$(BO$_{3}$)$_{2}$ in the field range 27 - 31~T covering the upper
phase boundary of the 1/8 magnetization plateau, identified at 28.4~T. 
Our data provide a clear evidence that above 28.4~T the spin-superlattice 
of the 1/8 plateau is modified but does not melt even though the magnetization 
increases. Although this is precisely what is expected for a supersolid phase,
the microscopic nature of this new phase is much more complex.  We discuss the 
field-temperature phase diagram on the basis of our NMR data.
\end{abstract}

\pacs{75.10.Jm, 75.25.+z, 75.30.Kz, 76.60.Jx}

\maketitle

A variety of novel quantum phenomena has been discovered by applying a magnetic
field to interacting systems of $S$=1/2 spin dimers, which have a singlet 
ground state and an energy gap $\Delta$ to the triplet excitations at zero field
\cite{rice021,nikuni001,kageyama991,kodama022,clemancey071,sebastian061}. 
A magnetic field splits the triplets and closes the gap at the critical field 
$H_{c}$=$\Delta/g\mu_{B}$.  Above $H_{c}$, a finite density of triplets (magnetization) 
generally undergoes the Bose condensation, resulting in an antiferromagnetic order 
perpendicular to the field \cite{affleck911,Giamarchi991,nikuni001,matsumoto041}.  Another remarkable 
phenomenon that may occur at higher fields is the magnetization plateau, where the 
magnetization stays constant at a fractional value of the saturation over a finite range of 
magnetic field \cite{kageyama991,onizuka001,ono031}.  In particular, a plateau with 
non-integer density of triplets per unit cell generally implies localization of the triplets into a 
spin-superlattice breaking the translational symmetry of the crystal \cite{oshikawa971}.  
Such a state is stabilized when the kinetic energy of the triplets is reduced, e.g. by 
frustration, so that the repulsive interaction becomes dominant.  

The nature of phases between the plateaux is particularly interesting, since
they are the bosonic analog of doped Mott insulators.  A simple scenario for an
incommensurate density of triplets is the melting of the superlattice and the
appearance of Bose condensation.  A more exotic possibility predicted for certain
dimer spin models \cite{ng061,laflorencie071,sengupta071,schmidt071} is the 
formation of a supersolid phase, where the interstitial triplets undergo Bose 
condensation in the background of the commensurate superlattice.  This means 
that the longitudinal spin density modulation coexists with a new transverse staggered
order.

\begin{figure}[b]
\includegraphics[width=1\linewidth]{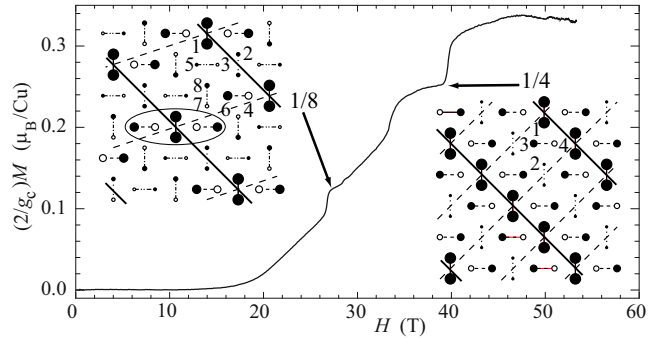}
\caption{\label{fig:magnetization} The magnetization divided by $g_{c}/2$
($g_{c}$=2.28 \cite{nojiri991}) is plotted against the pulsed magnetic field
along the $c$-axis (taken from Ref.~\cite{onizuka001}).  The measurement
was performed at $T$=1.4~K.  However, since the 1/8 plateau is not observed
above 0.6~K in steady fields \cite{takigawa041,levy}, we believe that the sample 
was adiabatically cooled during application of the pulse field. 
Also shown are the calculated spin density profiles for the 1/8 and 1/4
plateaux \cite{miyahara032}.  The solid (open) circles indicate magnetization
of Cu ions parallel (antiparallel) to the external field with the magnitude
represented by the circle size. The ellipsoid indicates a triplet-unit extending over
three dimers.}
\end{figure}
To date only a few spin systems are known to show symmetry-breaking magnetization 
plateaus. The best example is SrCu$_{2}$(BO$_{3}$)$_{2}$ \cite{kageyama991}, an orthogonal 
dimer system of Cu$^{2+}$ ions ($S$=1/2) with the primary interactions described by the 
two-dimensional Shastry-Sutherland model \cite{shastry811,miyahara031}.  This material 
shows a gap $\Delta$=35~K to the triplet excitations, which have a very small dispersion width 
(kinetic energy) \cite{kageyama001,gaulin041}, and plateaux at 1/8, 1/4 and 1/3 of the saturated
magnetization (see Fig.~1) \cite{onizuka001}.  A symmetry breaking spin-superlattice in the 
1/8 plateau has been confirmed by NMR experiments \cite{kodama022,takigawa041}.  
Furthermore, a supersolid phase has been proposed for the Shastry-Sutherland model 
above the 1/3 plateau \cite{momoi001}. However, SrCu$_{2}$(BO$_{3}$)$_{2}$ also contains 
an intradimer Dzyaloshinski-Moriya (DM) interaction and a staggered $g$-tensor, which  
break the spin-rotation symmetry and prevent true Bose condensation \cite{kodama051,zorko041,note1}. 
This explains the gradual increase of magnetization starting far below the expected critical field 
$\Delta/g\mu_{B}$=23~T and the absence of phase transitions up to the boundary 
of the 1/8 plateau  (26.5~T).  Since anisotropic interactions are often present in real 
materials, understanding their influence on the potential supersolid phase is an important
issue. For SrCu$_{2}$(BO$_{3}$)$_{2}$, steady fields above the 1/3 plateau are yet 
unavailable and the magnetization shows a direct jump from 1/4 to 1/3 plateaux (Fig.~1).  
Therefore, the vicinity of the 1/8 plateau provides a unique opportunity to look for such novel 
phenomena.   

In this paper, we report results of nuclear magnetic resonance (NMR) experiments 
on boron sites in SrCu$_{2}$(BO$_{3}$)$_{2}$ in the field range 27 - 31~T. We 
identified the upper boundary of the 1/8 plateau phase at 28.4~T from a clear splitting
of NMR lines.  The range of the internal field distribution, however, is nearly unchanged 
across the boundary. This provides direct evidence that spatial order of highly polarized 
triplet dimers persists above the 1/8 plateau even though the magnetization changes 
continuously, pointing to an analog of the supersolid phase. In this study we used 
the same single crystal as was employed in our previous high field NMR 
experiments \cite{kodama022,takigawa041,kodama051}. 
The NMR measurements were performed using a dilution refrigerator installed in 
a 20~MW resistive magnet at the Grenoble High Magnetic Field Laboratory.  
The $c$-axis of the crystal was aligned along the field direction in-situ within 
one degree by observing the angular variation of the NMR spectra at 18~T.

\begin{figure}[b]
\includegraphics[width=1\linewidth]{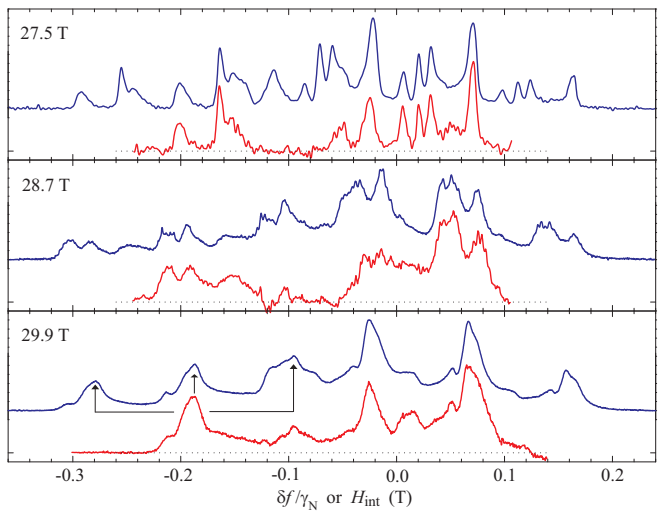}
\caption{\label{fig:spectra} (Color online) The upper blue lines show the $^{11}$B 
NMR spectra at 27.5~T (inside the 1/8 plateau) and at 28.7 and 29.9~T (outside the plateau) 
obtained at $T$=0.19~K.  The lower red lines show the distribution of the internal field 
$H_{\rm int}$ (see the text). The arrows in the lowest panel show an example of the 
quadrupole split three peaks deconvoluted into a single peak. The global similarity of 
all three spectra confirms the presence of spin-superlattice above the 1/8 plateau.}
\end{figure}
Figure 2 shows the $^{11}$B NMR spectra at $T$=0.19~K obtained at three
different field values, $H_{\rm ext}$=27.5, 28.7 and 29.9~T. The upper blue
lines represent the distribution of the frequency shift divided by the nuclear
gyromagnetic ratio, $\delta f / \gamma_{N}$, where 
$\delta f = f - \gamma_{N} H_{\rm ext}$, $\gamma_{N}$=13.66~MHz/T 
and $f$ is the resonance frequency. The spectrum taken at 27.5~T is representative 
for the entire 1/8 plateau as its shape does not change in the whole field range 
26.8 - 28.2~T \cite{kodama022} (data not shown).  The distinct spectral shapes 
at 28.7 and 29.9~T indicate that these fields are outside the plateau.  Note that the 
$^{11}$B NMR spectrum is determined by the distribution of the internal magnetic field 
$H_{\rm int}$ at nuclear sites, produced by the field-induced spin density, as well as by 
the quadrupole splitting $\nu_{Q}$ for the spin 3/2 nuclei
\begin{equation}
\delta f_{i} / \gamma_{N} = H_{\rm int} + i \nu_{Q}/\gamma_{N} ;~~i = -1, 0, 1.
\end{equation}
The spectrum thus represents convolution of the distribution function of
$H_{\rm int}$ with the sum of three delta functions separated by  
$\nu_{Q}/\gamma_{N}$.  One can recognize the three-fold periodic structure for all 
the spectra shown in Fig.~2 (an example is indicated by the arrows in the lowest panel) with 
$\nu_{Q}$=1.25~MHz.  This value is the same as obtained in low fields \cite{kodama021}, 
indicating that magnetic field does not induce strong lattice distortion. The distribution of 
$H_{\rm int}$ can then be obtained by deconvoluting the spectra using 
the inverse Fourier transform (the lower red lines in Fig.~2).

The distribution of $H_{\rm int}$ at 27.5~T agrees with the previous results obtained 
in a more primitive way \cite{takigawa041}.  It was shown in ref.~\cite{takigawa041} 
that the many sharp peaks distributed over a wide range of $H_{\rm int}$ are well 
reproduced by the spin density distribution of the Shastry-Sutherland model calculated 
by exact diagonalization \cite{kodama022,miyahara032}, which predicted a superlattice 
of highly polarized triplets surrounded by oscillating spin density shown in Fig.~1. 
In particular, the leftmost peak with the largest absolute value of the internal field, 
$H_{\rm int}= -0.2$~T, is assigned to the B site nearest to the Cu site bearing 
the largest magnetization (site 1 in Fig.~1) \cite{takigawa041}. Such distribution 
of $H_{\rm int}$ disappears completely when the field is decreased below 26.5~T,  
which is the lower boundary of the 1/8 plateau \cite{kodama022}.  In particular, 
there is no other ordered phase corresponding to a smaller fraction of
magnetization such as 1/9, in contradiction to Ref. \cite{sebastian}. 
The situation is very different when 
the field is increased \textit{above} the plateau.  The overall width of the distribution 
of $H_{\rm int}$ in Fig.~2 remains nearly unchanged. This is the direct evidence 
that \textit{highly polarized triplets are still present in the high field phase} even 
though the magnetization deviates from a commensurate fraction. However, 
unlike in the 1/8 plateau, the spectral shape changes in the high field phase, 
indicating some evolution of the spin structure. 

\begin{figure}[t]
\includegraphics[width=1\linewidth]{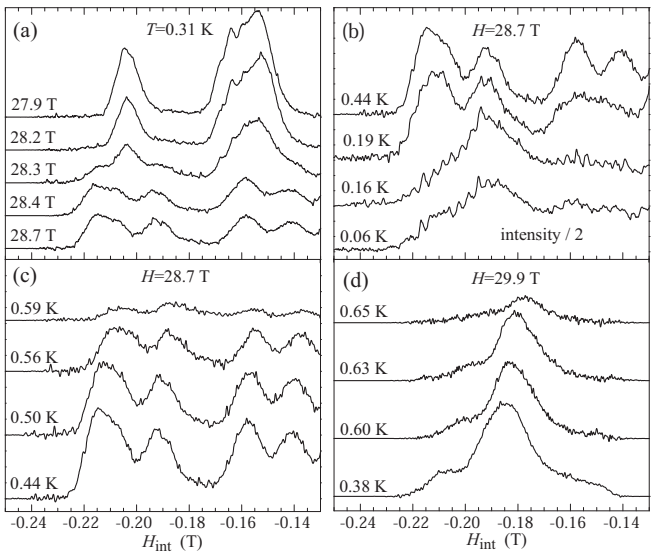}
\caption{\label{fig:transition}Variation of the low frequency part of the
$^{11}$B NMR spectra across the phase boundaries. See the text for details.}
\end{figure}
We now examine how the spectrum changes in more detail.  
Figure 3 shows the low frequency part of the spectrum corresponding to 
$H_{\rm int} \le -0.13$~T with changing field or 
temperature: (a) With increasing field at $T$=0.31~K,  the two most negative 
peaks get split above $H_{\rm ext}$=28.4~T, marking the upper boundary 
of the 1/8 plateau.  The presence of the split and unsplit peaks
at 28.3~T indicates coexistence of the two phases, pointing to a first order
transition. The transition field remains the same at $T$=0.12~K (data not
shown).  (b) With decreasing temperature at $H_{\rm ext}$=28.7~T, i.e. slightly
above the plateau boundary, one of the split peaks disappears suddenly between
0.19 and 0.16~K, suggesting another phase boundary. (c) With increasing
temperature at  $H_{\rm ext}$=28.7~T, the intensity rapidly decreases and 
the peaks completely disappear at $T_{c}$=0.60~K (zero intensity data are
not shown), marking the transition to the paramagnetic uniform phase. Indeed,
at higher temperatures only three quadrupole split NMR lines are observed
corresponding to a small \textit{uniform} internal field of $-0.03$~T. (d) With increasing
field, $T_{c}$ becomes higher: $T_{c}$=0.66~K at $H_{\rm ext}$=29.9~T and
$T_{c}$=0.71~K at $H_{\rm ext}$=30.9~T (data not shown). We notice that while 
the intensity of the peaks is reduced near $T_{c}$, their positions do not change 
significantly.  This means that the amplitude of the spin density modulation is 
only slightly reduced but the volume fraction of the ordered phase vanishes towards $T_{c}$, 
indicating again a first order transition.

\begin{figure}[b]
\includegraphics[width=1\linewidth]{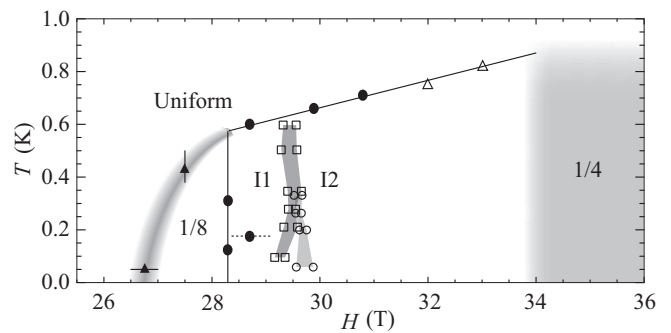}
\caption{\label{fig:phase} The phase diagram constructed from the present (solid circles) and the 
previous (solid triangles \cite{takigawa041}) NMR results and the specific 
heat data (open triangles \cite{tsujii031}). The bars indicate the width of the 
transition.  The transitions near 29.5~T detected by the torque measurements \cite{levy} 
are shown by open circles (increasing field) and open squares (decreasing field).  
The two data points at the same temperature represent the width of the transition. 
The lines and the shades are guide to the eyes.}
\end{figure}
The phase boundaries thus determined are plotted in Fig.~4, together with
the previous NMR data on the transition between the 1/8 plateau and the uniform
phase \cite{takigawa041}.  Our $H$-$T$ phase diagram is in good agreement 
with the recent magnetization and torque measurements in steady magnetic 
fields \cite{levy}, which revealed an abrupt change without hysteresis upon 
entering into the 1/8 plateau and a jump with hysteresis upon leaving the 1/8 plateau.  
In addition, another first order transition with hysteresis near 29.5~T was observed
as shown in Fig.~4, which was only weakly detected by NMR (data not shown.  
Implicitly, this is clear from the difference between the spectra of Figs.~3c and 3d). 
Thus there are at least two distinct intermediate phases (I1 and I2 in Fig.~4) between the 
1/8 and the 1/4 plateaux.  Our $T_{c}$ data extrapolate smoothly to the 
peak temperatures of the specific heat at higher fields \cite{tsujii031} also 
shown in Fig.~4.  From our NMR results it is now clear that the primary order 
parameter is the superlattice modulation of the longitudinal magnetization $\langle S_{z}
\rangle$.  In the intermediate phases, $T_{c}$ increases linearly with field.  Such 
behavior is distinct from the \textit{dome}-like variation in a typical field-induced 
Bose condensed phase \cite{sebastian061}. The transition at $T_{c}$ occurs 
in a narrow temperature window ($\leq$ 20~mK) in contrast to the broad width 
(more than 100~mK) observed for the transition between the 1/8 plateau and the 
uniform phase \cite{takigawa041}.

What is the nature of the intermediate phases ? Our main conclusion is that 
in both I1 and I2 phases, triplet dimers with large static magnetization are still 
present, i.e. a spin-superlattice persists.  The similarity of the internal field distribution 
for different fields in Fig.~2 suggests that the spin structure is locally similar to 
the one in the 1/8 plateau.  In particular, the \textit{triple-dimer} structure 
of the triplet unit consisting of the central dimer with parallel moments and the two 
neighboring orthogonal dimers with staggered moments (the ellipsoid in Fig.~1) 
should be preserved.  

The clear line splitting at the transition from the 1/8 plateau to the I1 phase (Fig.~3a) suggests
additional symmetry breaking, as expect for a supersolid phase.  Indeed, the kinetic 
energy of triplets at high fields is enhanced due to the correlated hopping process in 
the Shastry-Sutherland model \cite{momoi001}, and this should promote the formation 
of a supersolid phase \cite{schmidt071}.  However, as mentioned earlier, the DM interaction and
the staggered $g$-tensor in SrCu$_{2}$(BO$_{3}$)$_{2}$ break the continuous spin 
rotation symmetry in the $ab$-plane, which is prerequisite for a true supersolid transition.  
Nevertheless, it is possible that the transition causes breaking of some remaining discrete 
symmetries.  The anisotropic interactions and the resultant transverse 
staggered magnetization \cite{kodama051} obey all the discrete symmetries of the crystal 
including the $C_{2}$ rotation around the $c$-axis.  We note that the spin superstructure 
of the 1/8 plateau shown in Fig.~1 is also invariant under the $C_{2}$ rotation at the center 
of the triplet dimer, giving the same value of $H_{\rm int}$ for the two B nuclei nearest 
to the Cu at site 1. However, if a new long range order of the transverse magnetization 
breaking the $C_{2}$ symmetry occurs in the I1 phase, the NMR line from these B sites 
should split as observed in Fig.~3a \cite{note2}.  This would be analogous to a supersolid 
phase since the diagonal long range order (superlattice of $\langle S_{z} \rangle$) coexists 
with some kind of off-diagonal long range order (transverse staggered magnetization).  

In the I2 phase, the spectrum gets broadened and somewhat featureless, suggesting 
increased disorder or an incommensurate structure. We note that the stripe structure predicted
for the 1/4 plateau can be obtained by doping triple-dimer triplet units into the vacant 
sites in the rhomboid cell of the 1/8 plateau (Fig.~1). This will maintain similar local spin 
structure but introduce certain disorder if the doped triplets localize randomly. The same 
process is possible by starting from the square cell of the 1/8 plateau, which is obtained 
by shifting one of the two stripes of the rhomboid cell by half of its period.  Since both 
supercells are energetically nearly degenerate \cite{miyahara032}, they may alternate in the 
I1 phase, doubling the unit cell. This could be another explanation for the line splitting at the 
upper boundary of the 1/8 plateau and may lead to an incommensurate structure at higher doping.

Finally, we mention possible relevance of the theory describing effects of the DM
interaction on the frustrated spin ladders exhibiting the half magnetization plateau \cite{penc071}.  
It was found that the DM interaction, if it competes with the interdimer exchange, stabilizes the 
superlattice modulation of $\langle S_{z} \rangle$ far beyond the original plateau region in the 
phase diagram, similar to what occurs in SrCu$_{2}$(BO$_{3}$)$_{2}$.  In this model, however, 
the DM interaction also transforms the phase transitions separating the plateau and the neighboring 
phases into a crossover, in contrast to our observation in SrCu$_{2}$(BO$_{3}$)$_{2}$.  This is because 
the DM interaction mixes the singlet and the triplet states, hence $\langle S_{z} \rangle$ no longer 
represents the conserved boson density.  To clarify the effects of the DM interaction in more 
complicated Shastry-Sutherland model would be a future theoretical challenge.
 
In conclusion, our NMR results in SrCu$_{2}$(BO$_{3}$)$_{2}$ demonstrate that a
superlattice of highly magnetized triplet dimers persists above the 1/8 magnetization 
plateau, where the magnetization is no longer constant.  Although the presence of the DM 
interaction prevents the formation of a genuine supersolid phase, the NMR line splitting at the 
upper boundary of the 1/8 plateau suggests a new ordered phase of different symmetry.  
The magnetic phase diagram between the 1/8 and the 1/4 plateau is much more complex 
than expected and requires further experimental and theoretical work.

We are grateful for stimulating discussions with F. Mila, S. Miyahara, O. Tchernyshyov
and C. D. Batista on theoretical aspects and with F. L\'{e}vy and  I. Sheikin on the
magnetization and torque data.  The work was supported by Grant-in-Aids for
Scientific Research on Priority Areas Nos. 16076204 and 19052004 from the 
MEXT Japan, and by the French ANR Grant No. 06-BLAN-0111.


\begin{thebibliography}{99}
\bibitem{note1} The magnitude $D$ of the DM interaction and the staggered components $g_{s}$ of the $g$ tensor 
are estimated as $D/J$=0.034 and  $g_{s}$=0.023, where $J$ is the intradimer exchange \cite{kodama051}.
\bibitem{note2} A similar situation with two types of transverse magnetization has been
observed in another dimer compound, Cu$_{2}$(C$_{5}$H$_{12}$N$_{2}$)$_{2}$C$_{14}$ (Cu(Hp)Cl) \cite{clemancey071}.
\end{thebibliography}
\end{document}